\documentclass[fleqn, usenatbib]{mnras}
\usepackage{amssymb,amsmath}
\usepackage{graphicx}
\usepackage{xcolor}
\usepackage{multirow}
\usepackage[T1]{fontenc}
\usepackage{ae,aecompl}
\usepackage{newtxtext,newtxmath}

\def\msig{$M_{\rm BH}-\sigma$\ }

\def\obj{SDSS J1052+1036}

\title[Very blue-shifted broad H$\alpha$]
{Very blue-shifted broad H$\alpha$ in a low redshift Type-1.9 AGN: a disk emitter or a recoiling black hole scenario}
\author[Zhang X. G.]
{Xue-Guang Zhang$^{1}$\thanks{Corresponding author Email: \href{mailto:xgzhang@gxu.edu.cn}{xgzhang@gxu.edu.cn}}\\
$^1$ Guangxi Key Laboratory for Relativistic Astrophysics, School of Physical Science and Technology, GuangXi University, 
530004, Nanning, P. R. China}% \\
\pubyear{2023}

\begin{document}

\label{firstpage}

\pagerange{\pageref{firstpage}--\pageref{lastpage}}

\maketitle

\begin{abstract} %%%about 196 words
	In this manuscript, very blue-shifted broad H$\alpha$ with shifted velocity $\sim$2200km/s is reported in the low redshift 
Type-1.9 AGN SDSS J1052+1036. Blue-shifted broad emission lines may arise due to the presence of a rotating gas disk around central 
black hole (BH), but may also be a signature of rare phenomena such as gravitational wave recoil of a supermassive BH (rSMBH) or 
the presence of a binary BH (BBH) system. Here, due to larger shifted velocity of stronger and wider blue-shifted broad H$\alpha$, 
the BBH system is disfavoured. Meanwhile, if this object contained a rSMBH, intrinsic obscuration with E(B-V)$\le$0.6 should lead 
to a detectable broad H$\beta$, indicating the rSMBH scenario not preferred. We find that the blue-shifted broad H$\alpha$ can be 
well explained by emission from an AGN disk, indicating that SDSS J1052+1036 is likely a disk-emitting AGN. In order to determine 
which scenario, a rSMBH or a disk emitter, is more preferred, a re-observed spectrum in 2025 can provide robust clues, with a disk 
emitter probably leading to clear variations of peak positions, peak separations and/or peak intensity ratios in broad H$\alpha$, 
but with a rSMBH scenario probably leading to no variations of peak separations in broad H$\alpha$.
\end{abstract}


\section{Introduction}

%%1-1
	Shifted broad emission lines relative to stellar absorption features (or narrow emission lines) may be indicators of a 
gravitational wave recoiling supermassive black hole (rSMBH) in called off-nucleus active galactic nuclei (AGN), due to 
gravitational wave carried off linear momentum leading central BH being kicked away from central region of AGN, as discussed in 
\citet{bj73, mq04, ms06, vm07, bl08, km08a, bs16}. Broad emission lines from broad emission line regions (BLRs) bound to a rSMBH 
with a large kick velocity can lead to blue-shifted broad emission lines relative to narrow emission lines of AGN, due to no 
effects of a rSMBH on NLRs (narrow emission line regions). Until now, there are a few individual AGN and samples of AGN reported 
with blue-shifted broad emission lines, and expected rSMBH scenarios have been discussed in the literature.

\begin{figure*}
\centering\includegraphics[width = 18cm,height=12cm]{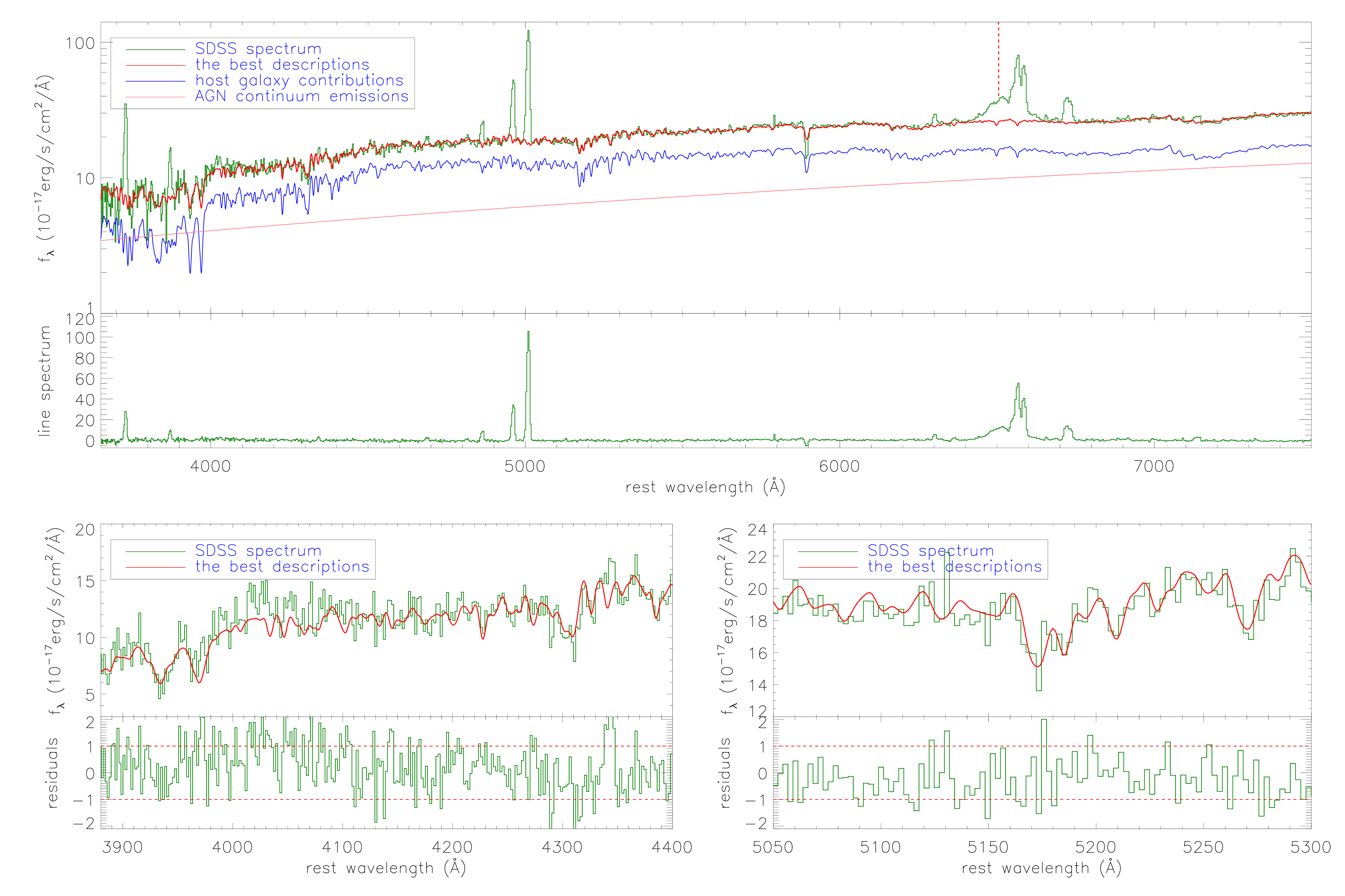}
\caption{Top region of the top panel shows the SSP method determined the best descriptions (solid red line) to the SDSS spectrum 
(solid dark green line) with emission lines being masked out. Solid blue line and solid pink line show the determined host galaxy 
contributions and the power law AGN continuum emissions, vertical dashed red line marks the blue-shifted broad H$\alpha$. Bottom 
region of the top panel shows the pure line spectrum calculated by the SDSS spectrum minus the sum of the host galaxy contributions 
and the power law AGN continuum emissions. Top regions and bottom regions of the bottom panels show the best fitting results (solid 
red line) and the corresponding residuals (the line spectrum minus the best fitting results and then divided by uncertainties of 
the SDSS spectrum) to the absorption features (solid dark green line) around the Ca~{\sc ii} H+K (bottom left panel) and around the 
Mg~{\sc i} (bottom right panel). In the bottom region of each bottom panel, horizontal red dashed lines show residuals=$\pm1$, 
respectively.
}
\label{spec}
\end{figure*}

%%1-2
	\citet{kz08} have reported SDSS J0927+2943 ($z\sim0.713$) with blue-shifted velocities 2650km/s in broad emission lines, 
to support a rSMBH scenario. However, \citet{be09} have discussed a binary BH (BBH) system with mass ratio 0.1 also leading to 
the shifted features in SDSS J0927+2943. \citet{sr09} have reported blue-shifted velocity 3500km/s in broad H$\beta$ in SDSS 
J1050, however, BLRs lying into central accretion disk (=disk emitter) would be preferred to explain the blue-shifted broad 
H$\beta$, rather than the rSMBH scenario. \citet{ss12} have reported very blue-shifted broad emission lines in SDSS J0956+5128, 
however, either an extreme disk emitter or a rSMBH is not the preferred scenario to explain all of the observed features, 
especially the different profiles between broad Balmer lines and broad Mg~{\sc ii} line. \citet{ky17} have discussed the rSMBH 
candidate of CXO J1015+6259 ($z\sim0.35$) with blue-shifted velocity 175km/s in broad emission lines. \citet{ks17} have shown 
a rSMBH is one proposed scenario to explain the three strong emission-line nuclei with velocity offset 250km/s in SDSS J1056+5516 
($z\sim0.256$), as well as a triple BH accreting system. \citet{ky18} have applied an oscillating rSMBH scenario to explain the 
broad emission line variability properties in Mrk1018. \citet{ce17, ce18, mc22} have shown that the quasar 3C186 ($z\sim1.07$) 
have blue-shifted velocity 2140km/s in broad emission lines, consistent with expected results by a rSMBH.

%%1-3
	Meanwhile, there are samples of AGN with blue-shifted broad emission lines. \citet{eb12} and followed in \citet{re15, 
re17} have reported a sample of tens of low redshift (z<0.7) SDSS quasars with blue-shifted velocities larger than 1000km/s 
in broad H$\beta$, and discussed that BBH systems should be preferred in a fraction of the candidates, after carefully checked 
changes of peak velocities through multi-epoch spectra. \citet{lr14} have shown 10 rSMBH candidates in nearby galaxies with 
small displacements between central activity region and center of galaxy. \citet{ke16} have reported a sample of candidates 
with mean blue-shifted velocity about 150km/s for rSMBHs in SDSS quasars with redshift less than 0.25. \citet{wg21} have shown 
nine AGN that may be spatially offset from their host galaxies and are considered as candidates for rSMBHs.

%%1-4
	Based on the reported candidates of AGN with blue-shifted broad emission lines, besides the rSMBH scenarios, either the 
BBH or disk emitter hypotheses can be applied. Moreover, as discussed in \citet{km08b, sh19}, candidates of rSMBHs with large 
recoiling velocities at low redshift are extremely rare. Here, a candidate at redshift 0.088 is reported with blue-shifted velocity 
$\sim$2200km/s in broad H$\alpha$ in a Type-1.9 AGN SDSS J105232.97+103620.08 (=\obj), with different scenarios discussed. This manuscript is organized as follows. Section 2 presents the spectroscopic results of the Type-1.9 AGN \obj. Section 3 gives main 
discussions. Section 4 gives our final conclusions. And the cosmological parameters have been adopted as 
$H_{0}=70{\rm km\cdot s}^{-1}{\rm Mpc}^{-1}$, $\Omega_{\Lambda}=0.7$ and $\Omega_{\rm m}=0.3$.

%%%fig 2
\begin{figure*}
\centering\includegraphics[width = 18cm,height=16.5cm]{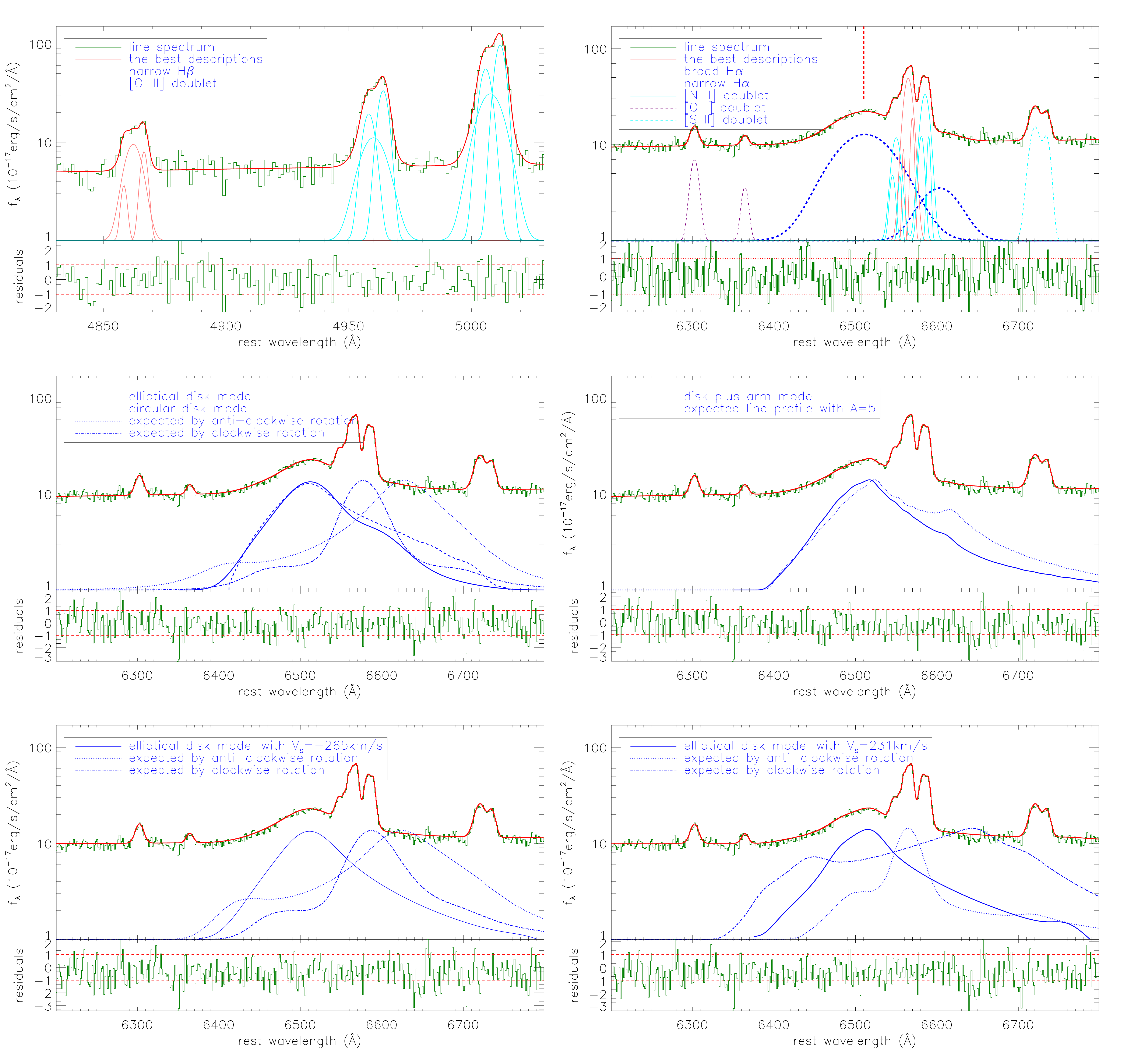}
\caption{Top left panel shows the best fitting results (solid red line) (top region) and the corresponding residuals (bottom 
region) to the emission lines around the H$\beta$ (solid dark green line) by multiple Gaussian functions. In top region of the 
top left panel, solid pink lines show the three Gaussian components in the narrow H$\beta$, solid cyan lines show the six Gaussian 
components in the [O~{\sc iii}] doublet. Top right panel shows the best fitting results (solid red line) (top region) and the 
corresponding residuals (bottom region) to the emission lines around the H$\alpha$ (solid dark green line) by multiple Gaussian 
functions. In top region of the top right panel, solid pink lines show the three Gaussian components in the narrow H$\alpha$, 
solid cyan lines shows the six Gaussian components in the [N~{\sc ii}] doublet, dashed purple lines and dashed cyan lines show 
the Gaussian components in the [O~{\sc i}] and [S~{\sc ii}] doublets, thick dashed blue lines show the two broad Gaussian components 
in the broad H$\alpha$. Middle left panel shows the best fitting results (solid red line) (top region) and the corresponding 
residuals (bottom region) to the emission lines around the H$\alpha$, with the elliptical accretion disk model applied to describe 
the broad H$\alpha$. In top region of the middle left panel, solid blue line shows the determined broad H$\alpha$ by the elliptical 
accretion disk model, dotted blue line and dot-dashed blue line represent the expected line profiles of the broad H$\alpha$ in 
Jul. 2025 with considering the standard elliptical accretion disk model applied with anti-clockwise rotation and clockwise 
processions respectively, dashed blue line shows the determined broad H$\alpha$ by a circular accretion disk model with $e=0$. 
Due to totally similar descriptions to the narrow emission lines, the Gaussian components are only shown in the top right panel 
for the narrow emission lines around the H$\alpha$. Middle right panel shows the best fitting results (solid red line) (top region) 
and the corresponding residuals (bottom region) to the emission lines around the H$\alpha$, with the circular accretion disk plus 
spiral arm model applied to describe the broad H$\alpha$. In top region of the middle right panel, solid blue line shows the model 
determined broad H$\alpha$, dotted blue line shows the expected line profile of the broad H$\alpha$ determined by the circular disk 
plus spiral arm model with different $A$. Bottom panels show the best fitting results (solid red line) (top regions) and the 
corresponding residuals (bottom regions) to the emission lines around H$\alpha$, with the elliptical accretion disk model with 
$V_s=-265$km/s (bottom left panel) and with $V_s=231$km/s (bottom right panel) applied to describe the broad H$\alpha$. In top 
regions of the bottom panels, dotted blue line and dot-dashed blue line represent the expected line profiles of broad H$\alpha$ 
in Jul. 2025 with considering the standard elliptical accretion disk model with $V_s=-265$km/s (bottom left panel) and with 
$V_s=231$km/s (bottom right panel) applied with anti-clockwise rotation and clockwise processions respectively. }
\label{line}
\end{figure*}

%%%%%Fig 3
\begin{figure*}
\centering\includegraphics[width = 18cm,height=13cm]{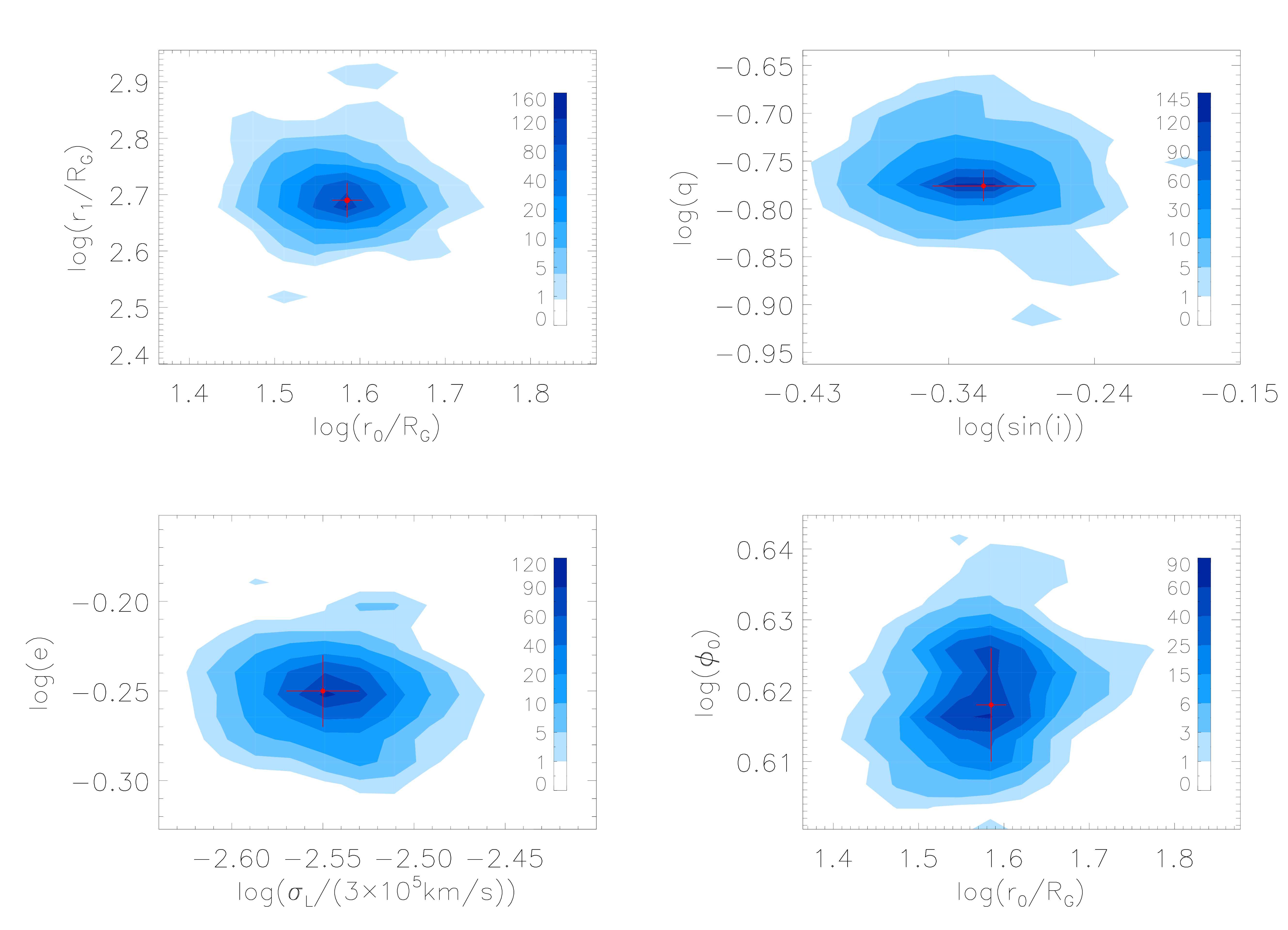}
\caption{MCMC technique determined two-dimensional posterior distributions in contour of the model parameters in the standard 
elliptical accretion disk model applied to describe the broad H$\alpha$. In each panel, sold circle plus error bars in red mark 
the positions of the accepted values and the corresponding $1\sigma$ uncertainties of the model parameters. The number densities 
related to different colors are shown in color bar in the left region of each panel.}
\label{mcmc}
\end{figure*}

\begin{figure*}
\centering\includegraphics[width = 18cm,height=18cm]{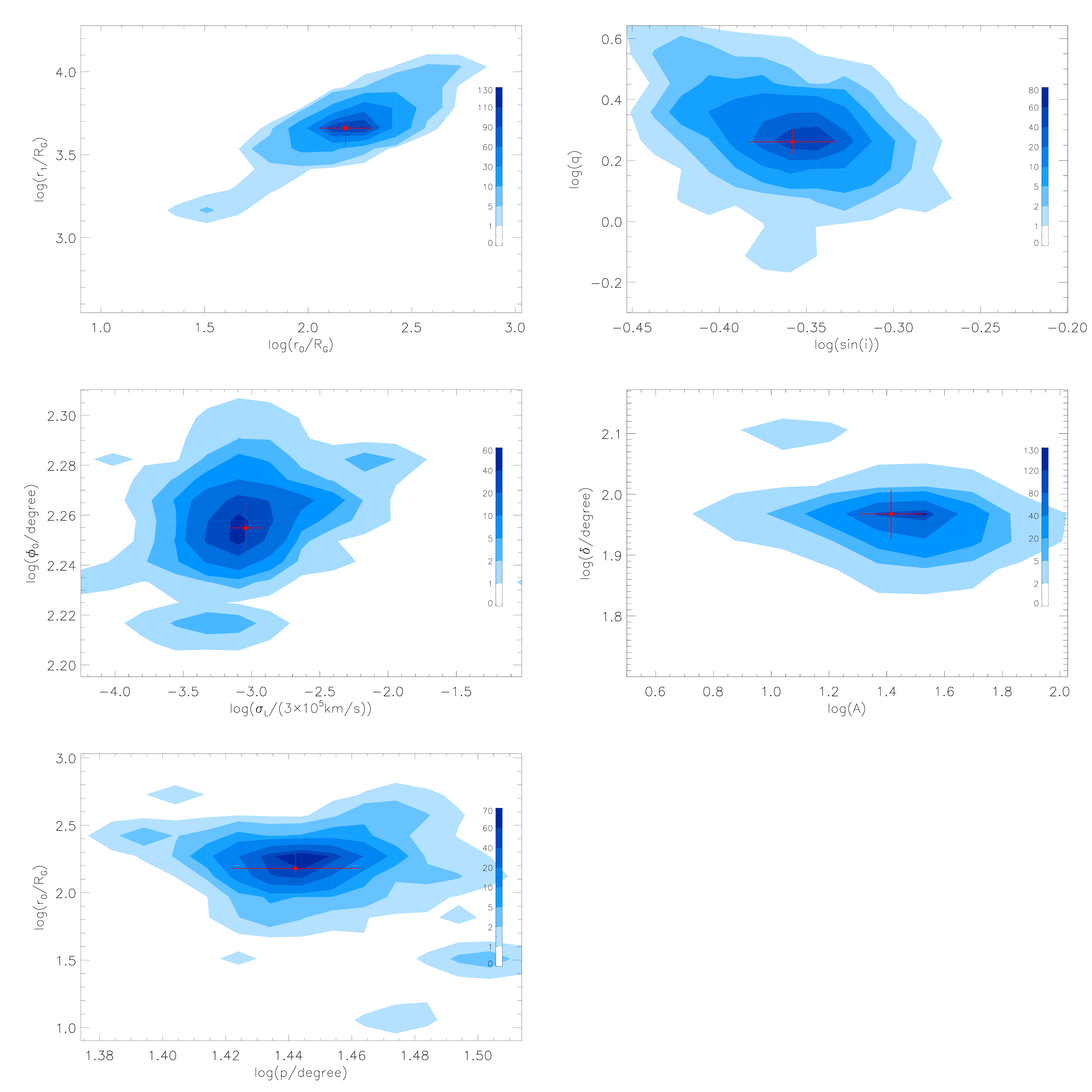}
\caption{MCMC technique determined two-dimensional posterior distributions in contour of the model parameters in the circular 
disk plus arm model applied to describe the broad H$\alpha$. In each panel, sold circle plus error bars in red mark the positions 
of the accepted values and the corresponding $1\sigma$ uncertainties of the model parameters. The number densities related to 
different colors are shown in color bar in the left region of each panel.}
	\label{arm}
\end{figure*}

\begin{figure*}
\centering\includegraphics[width = 18cm,height=5cm]{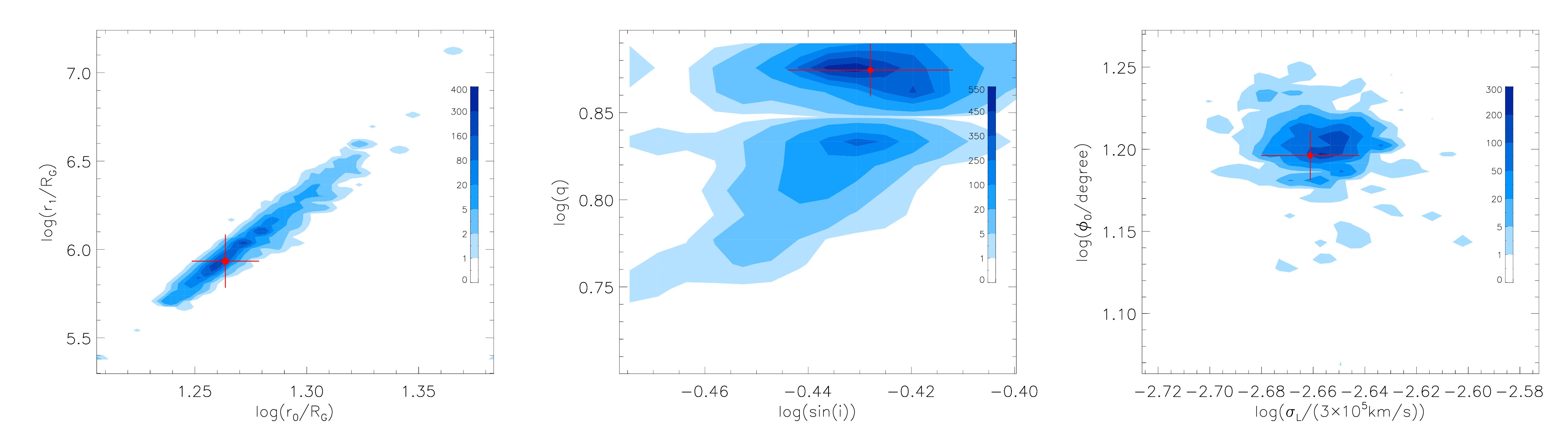}
\caption{MCMC technique determined two-dimensional posterior distributions in contour of the model parameters in the 
pure symmetric circular disk model applied to describe the broad H$\alpha$. In each panel, sold circle plus error bars in red 
mark the positions of the accepted values and the corresponding $1\sigma$ uncertainties of the model parameters. The number 
densities related to different colors are shown in color bar in the left region of each panel.}
\label{me0}
\end{figure*}

\section{Main results}% and discussions} 

%%2-1
	\obj~ is selected as the subject of this manuscript, due to its very blue-shifted broad H$\alpha$, while studying properties 
of double-peaked narrow emission lines in low redshift ($z<0.35$) SDSS quasars including some objects reported in the sample of 
\citet{ge12}. \obj~ has its SDSS spectrum (plate-mjd-fiberid=1602-53117-0243) with signal-to-noise about 18 shown in top left panel 
of Fig.~\ref{spec} with apparently shifted broad H$\alpha$ marked by vertical dashed red line.

%%2-2
	In order to measure the emission lines as well as to measure the stellar velocity dispersion, the commonly accepted SSP 
(Simple Stellar Population) method is applied to determine host galaxy contributions in \obj. More detailed descriptions on the 
SSP method can be found in \citet{bc03, ka03, cm05, cm17}. The SSP method has also been applied in our previous papers \citet{zh21a, 
zh21b, zh21m, zh22a, zh22b}. Here, we briefly describe the SSP method. The 39 simple stellar population templates from \citet{bc03, 
ka03} are applied to describe stellar lights, combined with a power law function to describe the AGN continuum. When the SSP method 
is applied, narrow emission lines are masked out by full width at zero intensity about 450km/s, and the spectrum with wavelength 
range from 6450\AA~ to 6750\AA~ are also masked out due to the strongly broad H$\alpha$. Then, through the Levenberg-Marquardt 
least-squares minimization technique (the MPFIT package), SDSS spectrum in rest frame with emission lines being masked out can 
be well described. The best descriptions and the corresponding line spectrum (SDSS spectrum minus the best descriptions) are shown 
in the top panel of Fig.~\ref{spec} with $\chi^2/dof\sim1.26$ (the summed squared residuals divided by degree of freedom). 
Considering the totally obscured broad H$\beta$, the determined red power law continuum emissions was acceptable, due to seriously 
obscurations on central continuum emissions. Meanwhile, we measured the stellar velocity dispersion to be 113$\pm$10km/s. Moreover, 
in order to show the stellar velocity dispersion, the bottom panels of Fig.~\ref{spec} show the SSP method determined the best 
descriptions and the corresponding residuals (SDSS spectrum minus the best descriptions and then divided by the uncertainties of 
SDSS spectrum) to the absorption features around Ca~{\sc ii} H+K from 3880 to 4400\AA~ and around Mg~{\sc i} from 5050 to 5300\AA.

%%2-3
	After subtractions of the host galaxy contributions, more apparent blue-shifted broad H$\alpha$ can be found. And the 
emission lines can be measured by multiple Gaussian functions, similar as what we have recently done in \citet{zh21a, zh21b, zh22a, 
zh22b, zh22c}. Considering the double-peaked features in the narrow emission lines (especially in the narrow Balmer lines, the 
[O~{\sc iii}] doublet and the [N~{\sc ii}] doublet) in \obj, three Gaussian functions are applied to describe each narrow emission 
line: two narrow Gaussian components for the double-peaked feature and one Gaussian component for the probably extended emissions 
underneath the double-peaked feature. Therefore, for the emission lines within the rest wavelength from 4830\AA~ to 5020\AA~ and 
from 6200\AA~ to 6800\AA, there are three Gaussian functions applied to describe the double-peaked narrow H$\beta$ (H$\alpha$), 
one broad Gaussian function to describe the probable broad H$\beta$, two broad Gaussian functions to describe the broad H$\alpha$, 
six Gaussian functions to describe the [O~{\sc iii}]$\lambda4959,5007$\AA~ doublet, six Gaussian functions to describe the 
[N~{\sc ii}]$\lambda6549,6585$\AA~ doublet, one Gaussian function to describe each line in the [O~{\sc i}]$\lambda6300,6363$\AA~ 
and the [S~{\sc ii}]$\lambda6716,6731$\AA~ doublets without apparent double-peaked features. When the functions above are applied, 
each Gaussian component has line intensity not smaller than zero, and the corresponding [O~{\sc iii}] ([N~{\sc ii}]) components 
have the same redshift and the same line width and have the flux ratio to be fixed to the theoretical value 3. 

%%2-4
	Then, through the MPFIT package, the best fitting results (in top regions) and the corresponding residuals (in bottom regions) 
to the emission lines around H$\beta$ and H$\alpha$ are shown in the top left panel and the top right panel of Fig.~\ref{line} with 
$\chi^2/dof\sim0.89$ and $\chi^2/dof\sim0.88$, respectively. Based on the best fitting results, it is not necessary to consider broad 
Gaussian component in the H$\beta$, because the determined line width and line flux (around to zero) of the broad Gaussian component 
are smaller than their corresponding uncertainties, indicating not apparent broad H$\beta$ in \obj. Moreover, as shown in the top right 
panel of Fig.~\ref{line}, each line in the [O~{\sc i}] and [S~{\sc ii}] doublets can be well described by one Gaussian component, and 
whether two or three Gaussian components applied to describe each line in the [O~{\sc i}] and [S~{\sc ii}] doublets have few effects 
on our discussed results on the broad H$\alpha$. The parameters of the Gaussian components applied to describe the emission lines 
are listed in the Table~1. Based on the best descriptions to the stellar absorption features in Fig.~\ref{spec} and the best fitting 
results to the broad H$\alpha$ in the top right panel of Fig.~\ref{line}, about 2200km/s blue-shifted broad H$\alpha$ can be confirmed.

%%2-5
	Besides the broad H$\alpha$ described by two Gaussian components, the blue-shifted broad H$\alpha$ can also be described 
by the known elliptical accretion disk model discussed in \citet{el95}, similar as what we have recently done on double-peaked 
broad emission lines in \citet{zh21c, zh22a}. The elliptical accretion disk model have seven model parameters, inner and out 
boundaries [$r_0$,~$r_1$] in unit of $R_G$ (Schwarzschild radius), inclination angle $i$ of disk-like BLRs, eccentricity $e$, 
orientation angle $\phi_0$ of elliptical rings, local broadening velocity $\sigma_L$ in units of ${\rm km/s}$, line emissivity 
slope $q$ ($f_r~\propto~r^{-q}$). In order to obtain more reliable model parameters and corresponding uncertainties, the Maximum 
Likelihood method combining with the MCMC (Markov Chain Monte Carlo) technique \citep{fh13} is applied. The evenly prior 
distributions of the seven model parameters are accepted with the following limitations, $\log(r_0)\in[1,~3]$,~$\log(r_1)\in[2,~6]$ 
($r_1~>~r_0$),~$\log(\sin(i))\in[-3,~0]$,~$\log(q)\in[-1,~1]$,~$\log(\sigma_L)\in[2,~4]$,~$\log(e)\in[-5,~0]$,
~$\log(\phi_0)\in[-5,~\log(2\times\pi)]$. The determined best fitting results and corresponding residuals to the emission lines 
around H$\alpha$ are shown in the top region and the bottom region of the middle left panel of Fig.~\ref{line} with 
$\chi^2/dof=324.41/391\sim0.83$. And the model determined best descriptions to the broad H$\alpha$ are shown as solid blue line 
in the top region of the middle left panel of Fig.~\ref{line}. The MCMC technique determined posterior distributions of the model 
parameters in the elliptical accretion disk model are shown in Fig.~\ref{mcmc}. The determined parameters and the corresponding 
$1\sigma$ uncertainties are listed in the Table~2. Moreover, as discussed in \citet{zh22a}, clean double-peaked broad line emission 
features can lead to solely determined model parameters in the elliptical accretion disk model. Therefore, there are no further 
discussions on whether is there solely determined model parameters, due to the apparent blue peak in broad H$\alpha$ in \obj. 

%%2-6
	Meanwhile, as suggested in \citet{el09, st03, st17}, rather than the elliptical accretion disk model, the spiral arm 
rotation is the preferred explanation for most disk emitter profile evolution. Therefore, the circular disk plus spiral arm 
model with 10 model parameters is also applied to describe the shifted broad H$\alpha$ of \obj. Besides the model parameters 
($e=0$) applied in the elliptical accretion disk model, four additional model parameters are applied to describe structures of 
spiral arms, the azimuthal width $\delta$, the pitch angle $p$ and the innermost radius $r_m$ of the spiral arm, and the brightness 
contrast $A$ between the spiral arm and the underlying axisymmetric disk. Then, based on the new emissivity formula shown in 
Equation (2) in \citet{st03} and accepted $r_m~=~r_0$, the best descriptions and the corresponding residuals to the emission lines 
around the H$\alpha$ are shown in the middle right panel of Fig.~\ref{line} with $\chi^2/dof=287.88/389\sim0.74$. And the model 
determined best descriptions to the broad H$\alpha$ are shown as solid blue line in the top region of the middle right panel of 
Fig.~\ref{line}. The MCMC technique determined posterior distributions of the model parameters in the circular disk plus arm 
model are shown in Fig.~\ref{arm}. The determined model parameters and the corresponding $1\sigma$ uncertainties are also 
listed in the Table~2 for the circular disk plus arm model.

%%%table -1
\begin{table}
\caption{parameters of the emission line components described by Gaussian functions}
\begin{tabular}{lccc}
\hline\hline
	line & $\lambda_0$ & $\sigma$ & flux \\
	     & \AA     & \AA & ${\rm 10^{-17}~erg/s/cm^2}$ \\
\hline
\multirow{2}{*}{broad H$\alpha$} & 6511.48$\pm$1.06 & 35.12$\pm$1.08 & 1036.8$\pm$29.2 \\
                                 & 6603.21$\pm$11.56 & 22.72$\pm$7.09 & 144.6$\pm$66.1 \\
\hline
\multirow{3}{*}{Narrow H$\alpha$} & 6558.81$\pm$0.31 & 1.32$\pm$0.38 &  26.5$\pm$11.4 \\  
	& 6569.66$\pm$0.21 & 1.97$\pm$0.31 &  89.5$\pm$27.3 \\
	& 6564.69$\pm$0.43 & 5.21$\pm$0.18 & 626.3$\pm$42.2 \\
\hline
\multirow{3}{*}{Narrow H$\beta$}& 4858.38$\pm$0.68 & 1.18$\pm$0.83 &  7.8$\pm$8.4 \\
	& 4866.56$\pm$0.28 & 1.42$\pm$0.34 & 24.9$\pm$9.9 \\
	& 4862.18$\pm$0.38 & 3.37$\pm$0.75 & 72.1$\pm$19.7\\
\hline
\multirow{3}{*}{[O~{\sc iii}]$\lambda5007$\AA} & 5005.81$\pm$0.25 & 2.38$\pm$0.31 & 327.6$\pm$141.4 \\
	& 5011.71$\pm$0.11 & 2.13$\pm$0.11 & 515.9$\pm$74.2 \\
	& 5007.72$\pm$0.41 & 5.09$\pm$0.82 & 383.2$\pm$200.2\\
\hline
[O~{\sc i}]$\lambda6300$\AA & 6302.43$\pm$0.44 & 4.86$\pm$0.46 &  72.1$\pm$6.1 \\
\hline
[O~{\sc i}]$\lambda6363$\AA & 6364.26$\pm$0.87 & 4.05$\pm$0.89 &  26.3$\pm$5.1 \\
\hline
\multirow{3}{*}{[N~{\sc ii}]$\lambda6583$\AA} & 6580.86$\pm$1.27 & 3.18$\pm$0.73 & 90.1$\pm$68.1 \\
	& 6589.87$\pm$0.38 & 1.93$\pm$0.52 & 53.8$\pm$32.3 \\
	& 6585.35$\pm$0.44 & 5.22$\pm$0.18 & 422.1$\pm$105.3 \\
\hline
[S~{\sc ii}]$\lambda6716$\AA & 6720.47$\pm$0.57 & 6.51$\pm$0.49 &  232.6$\pm$17.7 \\
\hline
[S~{\sc ii}]$\lambda6731$\AA & 6734.97$\pm$0.58 & 4.34$\pm$0.45 & 108.6$\pm$15.6 \\
\hline\hline
\end{tabular}\\
Notice: For the Gaussian components, the first column shows which line is measured, the Second, 
third, fourth columns show the measured line parameters: center wavelength $\lambda_0$ in unit of \AA, 
line width (second moment) $\sigma$ in unit of \AA~ and line flux in unit of ${\rm 10^{-17}~erg/s/cm^2}$. 
\end{table}

\begin{table}
\caption{model parameters of accretion disk models for the broad H$\alpha$}
\begin{tabular}{lccc}
\hline\hline
\multicolumn{4}{c}{the elliptical accretion disk model} \\
	\multicolumn{4}{c}{$r_0=38\pm2$,~$r_1=490\pm35$,~$\sin(i)=0.48\pm0.04$} \\
	\multicolumn{4}{c}{$q=0.17\pm0.01$,~$e=0.56\pm0.03$,~$\sigma_L=850\pm40{\rm km/s}$}\\
	\multicolumn{4}{c}{$\phi_0=237\pm5\degr$} \\
\hline\hline
	\multicolumn{4}{c}{the circular disk plus spiral arm model} \\
	\multicolumn{4}{c}{$r_0=150\pm60$,~$r_1=4500\pm1300$,~$\sin(i)=0.44\pm0.04$,~$q=1.82\pm0.21$} \\
	\multicolumn{4}{c}{$\sigma_L=270\pm90{\rm km/s}$,~$\phi_0=179\pm3\degr$, $A=26\pm7$}\\
	\multicolumn{4}{c}{$\delta=93\pm10\degr$, $p=28\pm2\degr$}\\
\hline\hline
	\multicolumn{4}{c}{the pure symmetric circular disk model} \\
	\multicolumn{4}{c}{$r_0=18\pm1$,~$r_1=(8\pm3)\times10^5$,~$\sin(i)=0.37\pm0.01$,~$q=7.48\pm0.91$} \\
	\multicolumn{4}{c}{$\sigma_L=650\pm103{\rm km/s}$,~$\phi_0=16\pm3\degr$}\\
\hline\hline
	\multicolumn{4}{c}{the elliptical accretion disk model with $V_s=-265$km/s} \\
	\multicolumn{4}{c}{$r_0=31\pm2$,~ $r_1=246\pm25$,~$\sin(i)=0.33\pm0.03$} \\
	\multicolumn{4}{c}{$q=0.66\pm0.05$,~$e=0.58\pm0.03$,~$\sigma_L=916\pm60{\rm km/s}$}\\
	\multicolumn{4}{c}{$\phi_0=240\pm6\degr$} \\
\hline\hline
\multicolumn{4}{c}{the elliptical accretion disk model with $V_s=231$km/s} \\
	\multicolumn{4}{c}{$r_0=44\pm4$,~ $r_1=381\pm40$,~$\sin(i)=0.38\pm0.04$} \\
	\multicolumn{4}{c}{$q=1.26\pm0.15$,~$e=0.54\pm0.03$,~$\sigma_L=532\pm30{\rm km/s}$}\\
	\multicolumn{4}{c}{$\phi_0=240\pm6\degr$} \\
	\hline\hline
\end{tabular}%\\
%Notice: For the Gaussian components, the first column shows which line is measured, the Second,
%third, fourth columns show the measured line parameters: center wavelength $\lambda_0$ in
%unit of \AA, line width (second moment) $\sigma$ in unit of \AA~ and line flux in unit
%of ${\rm 10^{-17}~erg/s/cm^2}$.
\end{table}

\section{main discussions}

\subsection{A kpc-scale dual core system in \obj?}

%%3.1-1
	Double-peaked [O~{\sc iii}]$\lambda5007$\AA~ can be seen in the spectrum of \obj, as well as shown in \citet{ge12}, 
widely indicating a kpc-scale dual core system \citep{zw04, xk09, fz11, wl19}. Based on the measured double-peaked features in 
the [O~{\sc iii}]$\lambda5007$\AA, the peak separation is about 350$\pm$22km/s in \obj, leading the broad emission lines from 
the assumed central two cores to have the same peak separation 350km/s. However, the peak separation about 4200$\pm$580km/s 
between the blue-shifted broad component and the red-shifted broad component in the broad H$\alpha$ in \obj~ is about twelve 
times higher than the peak separation of the double-peaked narrow emission lines. Therefore, the shifted broad H$\alpha$ is not 
related to a kpc-scale dual core system expected by the double-peaked [O~{\sc iii}] in \obj.

\subsection{A rSMBH in \obj?}

%%%3.2-1
        One another explanation for the blue shifted broad H$\alpha$ in \obj~ is that it is a rSMBH, after considering 
materials in the BLRs being carried away with the rSMBH. Meanwhile, not a single but two broad Gaussian components in the broad 
H$\alpha$ in \obj~ are probably indicating asymmetric structures of the BLRs bound to the rSMBH. As discussed in \citet{ms06, 
gm08, km08b}, the materials in the BLRs can be bound to a rSMBH within a region with the radius $r_k$ given by
\begin{equation}
r_k~\sim~512\frac{M_{BH}}{\rm 10^8M_\odot}(\frac{V_{k}}{\rm 10^3km/s})^{-2} {\rm light-days}
\end{equation} 
with $M_{BH}$ and $V_k$ as the BH mass and the kick velocity of a rSMBH. Meanwhile, in order to support a rSMBH by blue-shifted 
broad emission lines, the blue-shifted broad emission component related to the emission materials bound to a rSMBH should be 
apparent enough, indicating almost all the materials in the original BLRs bound to the rSMBH. Therefore, we can expect that 
the estimated $r_k$ should be not smaller than the origin BLRs size $R_{BLRs}$ which can be estimated by the continuum luminosity 
$L_{5100}$ at 5100\AA~ \citep{bd13},
\begin{equation}
\frac{r_k}{\rm light-days}~\ge~\frac{R_{BLRs}}{\rm light-days}~=10^{1.555+0.542\times\log(\frac{L_{5100}}{\rm 10^{44}erg/s})}
\end{equation}
In \obj~ with the well measured stellar velocity dispersion about 113$\pm$10km/s, after considering the M-sigma relation discussed 
in \citet{fm00, ge00, kh13, bb17, bt21} for both quiescent and active galaxies and also as discussed in \citet{ds05, jn09} in galaxy 
merging systems, the BH mass can be estimated as $2.5_{-1.3}^{+2.2}\times10^7{\rm M_\odot}$ in \obj, accepted Equation (7) in 
\citet{kh13}. Therefore, based on the equation above, we could have $L_{5100}~<~4\times10^{43}$erg/s, which can lead to apparent 
blue-shifted broad H$\alpha$ totally related to an expected rSMBH.

%%%3.2-2
	Based on the determined continuum emissions in the top panel of Fig.~\ref{spec}, the observed continuum luminosity at 
5100\AA~ is about $6.3\times10^{42}$erg/s in \obj. Accepted the intrinsic $L_{5100}$ should be smaller than $4\times10^{43}$erg/s, 
the intrinsic obscuration should have E(B-V)$\le$0.6. Then, accepted the intrinsic flux ratio 3.1 of broad H$\alpha$ to broad 
H$\beta$, the expected observed flux ratio of the broad H$\alpha$ to the broad H$\beta$ should be smaller than 6.2, leading to 
a detectable blue-shifted broad component in the H$\beta$ in \obj. Unfortunately, as shown in the top left panel of Fig.~\ref{line}, 
there are no detectable broad components in the H$\beta$ in \obj. Therefore, the blue-shifted broad H$\alpha$ probably contains 
weak contributions from a rSMBH scenario in \obj.

%%%3.2-3
	Unfortunately, the discussions above are not sufficient enough to totally disfavour the rSMBH scenario in \obj, however, 
multi-epoch spectroscopic results should provide clear clues to support or to be against the rSMBH scenario. If the expected 
rSMBH in \obj~ moves rectilinearly (or moves curvilinearly as the case in Mrk 1018\ in \citealt{ky18}), very tiny (or no) changes 
of peak separations between the blue-shifted component and the red-shifted component in the broad H$\alpha$ could be expected in \obj.

\subsection{A BBH system in \obj?}

%%%3.3-1
	If a BBH system was accepted in \obj~ with the estimated total BH mass $2.5_{-1.3}^{+2.2}\times10^7{\rm M_\odot}$ by 
the \msig relation, the two broad Gaussian components in the broad H$\alpha$ could be simply accepted to estimate the observational 
peak separation about $V_{p,obs}=4200\pm600{\rm km/s}$, leading the upper limit of the space separation $S$ of the central two BHs 
to be
\begin{equation}
S~<~\frac{G~\times~M_{BH}}{V_{p,obs}^2}~\sim~7.3_{-4.6}^{+11.3}{\rm light-days}
\end{equation}
Based on the measured luminosity $2.01\times10^{41}{\rm erg/s}$ of the observed broad H$\alpha$ or the measured continuum 
luminosity $6.2\times10^{42}{\rm erg/s}$ at 5100\AA~ in the rest frame, the estimated BLRs size should be about 7light-days, 
after considering the correlation between broad H$\alpha$ luminosity and continuum luminosity discussed in \citet{gh05} and 
the empirical R-L relation discussed in \citet{bd13}.

%%%3.3-2
	However, considering \obj~ as a Type-1.9 AGN, serious obscuration indicates the intrinsic BLRs size should be much 
larger than 7light-days. The BLRs size is similar as the upper limit of space separation of the central two BHs, strongly indicating 
the two BLRs probably totally mixed, leading to no apparent variability in the peak positions in the broad H$\alpha$, as discussed 
in \citet{sl10}. Moreover, under the assumption of a BBH system in \obj, probable optical quasi-periodic oscillations \citep{gd15a, 
gm15, zb16, zh22d, zh22e, zh23} should be detected. However, after checking long-term light curves from Catalina Sky Survey 
\citep{dd09}, All-Sky Automated Survey for Supernovae \citep{sp14, kc17} and Zwicky Transient Facility \citep{bk19, ds20}, there 
is no significant variability, which can not provide clues to support a BBH system in \obj.

%%%3.3-3
	Meanwhile, under the assumption of a BBH system in \obj, considering the stronger and wider broad blue-shifted component 
in the H$\alpha$ (H$\alpha_B$), the virial BH mass $M_{BH,B}$ related to the H$\alpha_B$ should be simply expected to be 6.4 times 
larger than the virial BH mass $M_{BH,R}$ related to the red-shifted broad component in the H$\alpha$ (H$\alpha_R$), accepted the 
virialization assumptions to the broad emission lines as discussed in \citet{gh05, pe04}. Here, the factor 6.4 is simply calculated 
by $(\frac{1036.8}{144.6})^{0.5}(\frac{35.12}{22.72})^2$ with 1036.8 and 144.6 (35.12 and 22.72) as the line fluxes (the line 
widths) of the H$\alpha_B$ and the H$\alpha_R$ in \obj. Then, the H$\alpha_B$ should have 6.3 times smaller shifted velocity than 
that of the H$\alpha_R$, which is against the measured results that the shifted velocity 2430$\pm$50km/s of the H$\alpha_B$ is 
larger than the shifted velocity about 1760$\pm$530km/s of the H$\alpha_R$, indicating a BBH system is disfavoured in \obj.

%%%3.3-4
	Furthermore, if we accepted the double-peaked narrow emission lines as signs of kpc-scale dual core systems and also 
accepted the blue-shifted broad H$\alpha$ related to a BBH system, there should be a rare close-pair binary in a triple BH system 
in \obj, similar as those discussed in \citet{hl07, dp14}. In such a rare close-pair binary in a triple BH system in \obj, 
similar results can be expected that the shifted velocity of the H$\alpha_B$ should be quite smaller than that of the H$\alpha_R$. 
However, whether the red-shifted (or the blue-shifted) narrow emission component in the double-peaked narrow H$\alpha$ is applied 
to trace the rotating velocity of the close-pair binary BH system in a triple BH system, larger shifted velocity of the 
H$\alpha_B$ can be determined than that of the H$\alpha_R$. Therefore, a close-pair binary in a triple BH system is disfavoured 
in \obj

\begin{figure*}
\centering\includegraphics[width = 12cm,height=8cm]{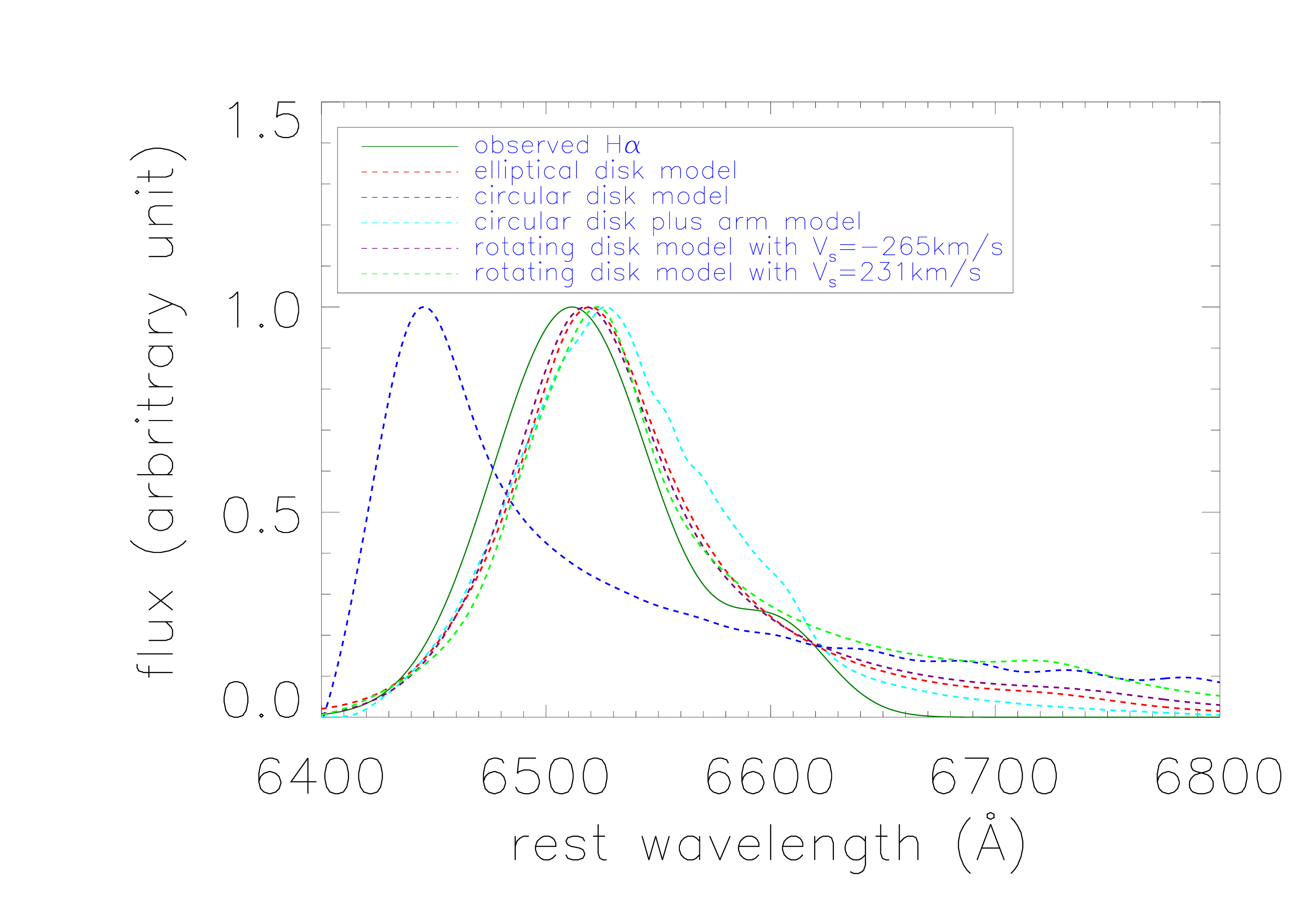}
\caption{Expected line profiles of the shifted broad H$\alpha$ with only $r_0$ and $r_1$ changed in the different 
accretion disk models. As described in the legend, besides the solid dark green line representing the observed 
shifted broad H$\alpha$ described by two broad Gaussian functions, each other dashed line in different color represents 
the corresponding model expected line profile of the shifted broad H$\alpha$ in different epoch with only $r_0$ and $r_1$ 
changed.}
\label{l3ne}
\end{figure*}

\subsection{A disk emitter in \obj?}

%%3.4 - 1
	Based on the model parameters of the elliptical accretion disk model listed in the Table~2, the expected disk precession 
period should be about $T_{\rm pre}\sim1040M_{8}R_{3}^{2.5}yr$. Using the \msig determined BH mass $M_{8}\sim0.25_{-0.13}^{+0.22}$ 
in units of $10^8{\rm M_\odot}$ and $R_{3}$ as radius in units of 1000$R_G$, based on the determined $r_0$, $r_1$ and $q$, the flux 
weighted size of the emission regions for the broad H$\alpha$ to the central BH is about 248$R_G$, leading to an approximately 
estimated disk precession period of 8years. As well known, asymmetric structures in accretion disk model are key factors leading 
to apparent variabilities of the peak positions and the peak separations of the double-peaked broad emission lines due to pure 
disk precessions, which will provide clues to support a disk emitter in \obj. If there should be a re-observed spectrum in Jul. 
2025 (MJD$\sim$60858), based on the expected precession period of about 8years, the expected line profiles of the broad H$\alpha$ 
in \obj~ in 2025 are shown as dotted blue line and dot-dashed blue line in the top region of the middle left panel of Fig.~\ref{line} 
with considering the standard elliptical accretion disk model applied with anti-clockwise rotation and clockwise processions 
respectively.

%%3.4 - 2
	Meanwhile, it is necessary to check whether a pure symmetric circular accretion disk model (with eccentricity to be 
zero) (without spiral arms) can be applied to describe the observed shifted broad H$\alpha$ in \obj. For a circular accretion 
disk model with $e=0$, a similar fitting procedure is applied to describe the broad H$\alpha$ in \obj, with the final determined 
fitting results to the broad H$\alpha$ shown as dashed blue line in the top region of the middle left panel of Fig.~\ref{line} 
with corresponding $\chi^2/dof=469.37/392\sim1.21$. The MCMC technique determined posterior distributions of the model parameters 
in the pure symmetric circular disk model are shown in Fig.~\ref{me0}. The determined model parameters and the corresponding 
$1\sigma$ uncertainties are also listed in the Table~2 for the pure symmetric circular disk model. Based on the F-test technique 
similar as what we have recently done in \citet{zh22c}, due to the different values of $\chi^2$ and $dof$ for the different 
accretion disk models, the confidence level can be determined to be higher than 6$\sigma$ to support that the elliptical accretion 
disk model and the circular accretion disk plus arm model is preferred than the pure symmetric circular accretion disk model. Unfortunately, only through the single-epoch spectroscopic properties of \obj, we can not find more clues to 
support that the pure symmetric circular disk model is totally disfavored in \obj. Therefore, in the manuscript, the pure symmetric 
circular disk model is also accepted as a reasonable model to describe the shifted broad H$\alpha$ in \obj.

%%3.4 - 3
	Moreover, if accepted the double-peaked features in the [O~{\sc iii}]$\lambda4959,5007$\AA~ doublet as signs of a kpc-scale 
dual core system, a rotating disk emitter (disk emission regions with a rotating velocity related to the orbital motions of central 
dual cores) contained in a dual core system could also be applied to describe the observed blue-shifted broad H$\alpha$ in \obj. 
Considering the double-peaked features in the narrow H$\alpha$ to trace the rotating velocity $V_s$ of the disk emitter in a dual 
core system, the best fitting results and the corresponding residuals to the emission lines around H$\alpha$ can be re-determined 
with $V_s=-265$km/s and with $V_s=231$km/s, and shown in the bottom left panel and the bottom right panel of Fig.~\ref{line} with 
corresponding $\chi^2/dof\sim1.03$ and $\chi^2/dof\sim1.02$, respectively. The model determined broad H$\alpha$ after considering 
$V_s$ are shown as solid blue lines in the top regions of the bottom panels of Fig.~\ref{line}. Moreover, if considering 
the elliptical accretion disk model with $V_s$ in \obj~, there are similar $1\sigma$ uncertainties of the model parameters as those 
of the standard elliptical accretion disk model shown in Fig.~\ref{mcmc}. Therefore, we did not show the posterior distributions of 
the model parameters for the rotating elliptical disk models, but the model parameters and the corresponding $1\sigma$ uncertainties 
are listed in the Table~2. Based on the determined model parameters listed in the Table~2 for the rotating elliptical accretion 
disk model with $V_s$, disk precession periods can be estimated as 1.68years and 4.95years, with the central wavelengths of the 
blue-shifted component and the red-shifted component in the narrow H$\alpha$ applied to determine the $V_s$. Then, the expected 
line profiles of the broad H$\alpha$ in \obj~ in Jul. 2025 are shown as dotted blue line and dot-dashed blue line in the top 
regions of the bottom left panel and the bottom right panel of Fig.~\ref{line} with considering the elliptical accretion disk model 
with $V_s=-265$km/s and $V_s=231$km/s applied with anti-clockwise rotation and clockwise processions respectively.

%%3.4 - 4
	Either a rotating elliptical disk emitter contained in a dual core system or a standard elliptical 
disk emitter can lead to apparent time dependent variations of the peak positions and the peak separations 
of the shifted broad H$\alpha$ in \obj~ as a disk emitter. Considering different disk precession periods 
determined by standard accretion disk model and/or rotating disk emitter, if there should be a re-observed 
spectrum in Jul. 2025, the expected broad H$\alpha$ in 2025\ in \obj~ have quite different peak positions 
and different peak separations from the broad H$\alpha$ in the SDSS spectrum observed in MJD=53117. Therefore, 
a re-observed spectrum in 2025 should provide clues enough to confirm whether a disk emitter is preferred 
in \obj. Unfortunately, unless there are detailed time-dependent variabilities of the broad H$\alpha$ in 
\obj, it is hard to distinguish a standard elliptical disk emitter from a rotating elliptical disk emitter 
in a kpc-scale dual core system.

%%3.4 - 5
	Furthermore, if considering the circular disk plus spiral arm model in \obj, the expected disk 
precession period should be about 2100years, due to the quite large flux weighted size $R_3\sim2.3$ of the 
emission regions of the blue-shifted broad H$\alpha$ to the central BH. Meanwhile, the precessing spiral 
pattern could have precession period more than 100years, considering ten times of the dynamical time of 
accretion disk as discussed in \citet{st03}. The large precession periods strongly indicates no apparent 
variabilities of profiles of the broad H$\alpha$ in \obj, if considering the circular disk plus arm model 
to describe the shifted broad H$\alpha$. However, as discussed in \citet{st03}, the model parameters of $A$, 
$q$ are varying in the circular disk plus spiral arm model. The varying parameter of $A$ can lead to apparent 
variability of the peak intensity ratio of the broad H$\alpha$ of \obj. As an example, if $A$ was changed 
from 26 to 5 in 2025, quite stronger red peak could be possibly expected in the broad H$\alpha$ of \obj, 
as shown in the top region of the middle right panel of Fig.~\ref{line}. Similar results can be found, if 
the circular plus arm model discussed in a dual core system. Therefore, there are no plots or discussions 
for the corresponding results based on the circular plus arm contained in a dual core system.

	Before the end of the subsection, an additional point should be noted. Discussions above on variations 
of the profiles of the shifted broad H$\alpha$ are due to pure disk precessions in \obj~ as a disk emitter, 
which can lead to apparent variations from the elliptical accretion disk model but no apparent variations 
from the circular disk model neither from the circular plus arm model. However, effects of variability of 
AGN activities should also have apparent effects on the variations of profiles of the shifted broad H$\alpha$ 
in \obj~ as a disk emitter, which will be discussed as follows.

	Sizes $R_{BLRs}$ of central BLRs of dozens of broad line AGN have been measured through the 
reverberation mapping technique \citep{bla82, pet93, pet99, bar15, du18, she19, luk23}, leading to the known 
dependence of $R_{BLRs}$ on AGN continuum luminosity (or on broad line luminosity) \citep{kas00, bd13}. 
Therefore, stronger central AGN emissions can lead to deeper ionization boundaries of central BLRs, in other 
words, stronger AGN emissions can lead to larger $R_{BLRs}$. Therefore, if there was apparent variability of 
AGN activities in \obj, different inner radius and outer radius in different epochs could be expected from 
those listed in the Table~2 of the different accretion disk models for shifted broad H$\alpha$ shown in 
Fig.~\ref{line}, leading to probably different line profiles. As an example, if we assumed that there was 
a spectrum observed again at any time but with the re-observed AGN continuum luminosity at 5100\AA~ about 2 
times of the AGN continuum luminosity shown in the top panel of Fig.~\ref{spec}, leading the expected 
$R_{BLRs}$ to be 1.4 times larger than the $R_{BLRs}$ for the shifted broad H$\alpha$ shown in Fig.~\ref{line}. 
Then, the re-determined inner radius and outer radius of the different disk models should be approximately 
1.4 times of the $r_0$ and $r_1$ listed in the Table~2. Based on the re-determined inner radius and outer 
radius and the other model parameters which remain unchanged, the model expected line profiles are shown in 
Fig.~\ref{l3ne} only considering AGN variability, leading to apparent variations of the line profiles of the 
shifted broad H$\alpha$ by different accretion disk models (even the pure symmetric circular disk model) 
in different epochs, strongly different from the expected results by the rSMBH scenario. Therefore, besides 
the accretion disk precessions, strong AGN variations have also apparent effects on variations of the profiles 
of the shifted broad H$\alpha$, which will provide further clues to support \obj~ as a disk emitter.

\section{Main Summary and Conclusions}

	A rare low-redshift Type-1.9 AGN \obj~ but with very blue-shifted broad H$\alpha$ is reported in this 
manuscript. The main summary and conclusions are as follows.
\begin{itemize}
\item The very blue-shifted broad H$\alpha$ is not related to a kpc-scale dual core system in \obj~ with 
	apparent double-peaked narrow emission lines, due to quite different peak separations of the 
	double-peaked narrow emission lines from the peak separation of the blue-shifted component and the 
	red-shifted component in the broad H$\alpha$.
\item The very blue-shifted broad H$\alpha$ is not related to a sub-pc BBH system in \obj, due to stronger and wider 
	blue-shifted H$\alpha$ having larger shifted velocity.
\item The very blue-shifted broad H$\alpha$ probably does not arise due to the rSMBH scenario, mainly due to 
	rSMBH scenario expected obscuration having E(B-V)$\le$0.6, leading to probably detectable broad 
	components in the H$\beta$, against the spectroscopic results without broad H$\beta$ in \obj. 
\item The very blue-shifted broad H$\alpha$ can be well explained by a disk emitter in \obj~ without any caveats.
\item Due to disk precessions of accretion disk models, the standard elliptical accretion disk model can lead to 
	apparent variations of the profiles of the shifted broad H$\alpha$ in \obj~ as a disk emitter, however 
	no apparent variations could be expected in recent years through the circular disk model or the circular 
	plus arm model. 
\item If considering strong variations of AGN activities leading to variations of ionization boundaries, 
	apparent variations in different epochs can be expected in the profiles of the shifted broad H$\alpha$ 
	in \obj~ as a disk emitter.
\item A re-observed spectrum in 2025 could provide robust clues to support a disk emitter in \obj, if there 
	were apparent variations of peak positions, peak separations and/or peak intensity ratios in the broad 
	H$\alpha$. 
\end{itemize}

\section*{Acknowledgements}
Zhang gratefully acknowledges the anonymous referee for giving us constructive comments and suggestions to 
greatly improve our paper. Zhang gratefully acknowledges the kind financial support from GuangXi University 
and the kind funding support NSFC-12173020 and NSFC-12373014. This research has made use of the data from the 
SDSS (\url{https://www.sdss.org/}) funded by the Alfred P. Sloan Foundation, the Participating Institutions, 
the National Science Foundation and the U.S. Department of Energy Office of Science. The research has made 
use of the MPFIT package \url{https://pages.physics.wisc.edu/~craigm/idl/cmpfit.html}, and of the emcee 
package \url{https://pypi.org/project/emcee/}.

\section*{Data Availability}
The data underlying this article will be shared on request to the corresponding author
(\href{mailto:xgzhang@gxu.edu.cn}{xgzhang@gxu.edu.cn}).

\label{lastpage}
\end{document}